\documentclass[copyright]{eptcs}
\pdfoutput=1
\providecommand{\event}{T. Neary, D. Woods, A.K. Seda and N. Murphy (Eds.):\\ The Complexity of Simple Programs 2008.}
\providecommand{\volume}{1}
\providecommand{\anno}{2009}
\providecommand{\firstpage}{81}
\usepackage{breakurl}

\usepackage{amsfonts,amssymb,amsthm,amsmath,url}
\usepackage{graphicx}

\newcommand{\setN}{\mathbb{N}}
\newcommand{\setZ}{\mathbb{Z}}

\newcommand\focus[1]{\par\textbf{#1}}

\newcommand{\ACA}{\mathcal{A}}
\newcommand{\ACB}{\mathcal{B}}

\newcommand{\globA}{G_{\ACA}}
\newcommand{\globB}{G_{\ACB}}

\newcommand\alphabe[1]{S_{#1}}
\newcommand{\alphA}{\alphabe{\ACA}}
\newcommand{\alphB}{\alphabe{\ACB}}
\newcommand\ZZ{\setZ}

\newcommand\bloc[1]{b_{#1}}
\newcommand\debloc[1]{b^{-1}_{#1}}
\newcommand\sac{\sqsubseteq}
\newcommand{\simu}{\preccurlyeq}
\newcommand\grp[2]{{#1}^{<#2>}}

\newcommand\moindre{\prec}

\newcommand\CC{\mathbf{cc}}
\newcommand\CCO{\mathbf{cc_1}}
\newcommand\CP{C_P}

\newtheorem{definition}{Definition}
\newtheorem{proposition}{Proposition}
\newtheorem{theorem}{Theorem}

\newtheorem{openproblem}{Open Problem}
\newtheorem*{remark}{Remark}

\newtheorem{corollary}{Corollary}
\newtheorem{fact}{Fact}
\newtheorem{notation}{Notation}

\newcommand{\singlerule}[4]{%
      \begin{tabular}[c]{c@{\hskip .2em}c@{\hskip .2em}c}
        &#1\\
        #2&#3&#4
      \end{tabular}
}

\newcommand{\carule}[8]{%
  \begin{center}
    \small
    \begin{tabular}[c]{cccccccc}
        \singlerule#1 000
&       \singlerule#2 001
&       \singlerule#3 010
&       \singlerule#4 011
&       \singlerule#5 100
&       \singlerule#6 101
&       \singlerule#7 110
&       \singlerule#8 111
    \end{tabular}%
  \end{center}
}

  \title{Communications in cellular automata\footnote{Partially
      supported by Programs Fondap, Basal-CMM, Fondecyt 1070022 (E.G),
      Fondecyt 1090156 (I.R.), and Instituto Milenio ICDB}}

  \author{Eric Goles%
    \institute{Facultad de Ingenier\'{\i}a y Ciencias, Universidad Adolfo
    Ib\'a\~nez, Santiago, Chile}
    \institute{Centro de Modelamiento Matem\'atico (UMI 2807 CNRS), 
   Universidad de Chile, Chile}
   \and
   Pierre-Etienne Meunier
  \institute{LAMA, Universit\'e de Savoie, CNRS, France}
    \and
  Ivan Rapaport
    \email{rapaport@dim.uchile.cl}
  \institute{Centro de Modelamiento Matem\'atico (UMI 2807 CNRS), 
   Universidad de Chile, Chile}
  \institute{Departamento de Ingenier\'{\i}a Matem\'atica, 
   Universidad de Chile, Chile}
  \and
  Guillaume Theyssier
  \institute{LAMA, Universit\'e de Savoie, CNRS, France}
  }
\def\titlerunning{Communications in cellular automata}
\def\authorrunning{E. Goles, P.-E. Meunier, I. Rapaport and G. Theyssier}
  
\begin{document}
\maketitle
%\begin{frontmatter}

\begin{abstract}
  The goal of this paper is to show why the framework of communication
complexity seems suitable for the study of cellular automata.
Researchers have tackled different algorithmic problems ranging from
the complexity of predicting to the decidability of different
dynamical properties of cellular automata. But the difference here is
that we look for communication protocols arising \emph{in the dynamics
itself}. Our work is guided by the following idea : \emph{if we are
able to give a protocol describing a cellular automaton, then we can
understand its behavior}.

\end{abstract}

%\begin{keyword}
%  % keywords here, in the form: keyword \sep keyword cellular
%  automata\sep communication complexity\sep classification\sep
%  universality.
%\end{keyword}
%\end{frontmatter}

\section{Cellular automata}
%% Definitions de base des automates cellulaires
%% elementaires. Peut-etre serait-il interessant de donner des
%% definitions plus generales, mais ceci dit, c'est suffisant pour
%% comprendre le reste.
Throughout this paper we restrict our study to one-dimensional
cellular automata. These are infinite collections of cells arranged
linearly, each having a state from a finite set. The dynamics of the
system is governed by a local rule applied uniformly and synchronously
to the lattice of cells.

A cellular automaton (CA) is a triple ${\ACA=(S, r, f)}$ where:
\begin{itemize}
\item $S$ is a (finite) \emph{state set},
\item $r$ is the neighborhood \emph{radius},
\item $f:S^{2r+1}\rightarrow S$ is the \emph{local transition function}.
\end{itemize}

A coloring of the lattice $\ZZ$ with states from $S$ (\textit{i.e.} an
element of $S^\ZZ$) is called a
\emph{configuration}.  To $\ACA$ we associate a global function
$G$ acting on configurations by synchronous and uniform
application of the local transition function.  Formally, $G:
S^\ZZ\rightarrow S^\ZZ$ is defined by:
\[G(x)_z = f(x_{z-r},\ldots,x_{z+r})\] for all ${z\in\ZZ}$.  Several
CA can share the same global function although there are
syntactically different (different radii and local
functions). However, as we will see below (section~\ref{sec:ccinca}),
the main property we are interested in (namely, communication
complexity) is independant of the particular choice of the syntactical
representation. Moreover, an important part of the paper
(Section~\ref{classes}) focuses on elementary CA, which is a fixed
syntactical framework.

After $n$ time steps the value of a cell depends on its own initial
state together with the initial states of the $rn$ left and $rn$ right
neighbouring cells.  More precisely, we define the $n$-th iteration of
local rule $f^n:\{0,1\}^{2rn+1} \rightarrow \{0,1\}$ recursively:
${f^1=f}$ and, for $n \geq 2$,
\[f^n(z_{-rn} \ldots z_1,z_0,z_1 \ldots z_{rn})=
f^{n-1}( f(z_{-rn},\ldots,z_{-rn+2r}) 
\ldots f(z_{rn-2r},\ldots,z_{rn}) ).\]

%\begin{definition}[Wolfram numbers]
%  There are as many elementary cellular automata as there are
%  functions of $\{0,1\}^{\{0,1\}^3}$, i.e. 256, and each of them is
%  identified with its Wolfram number $\omega=\sum_{a,b,c\in\{0,1\}}
%  2^{4a+2b+c}f(a,b,c)$ (see~\cite{wolfram84,wolfram02}).
%\end{definition}
%For instance, rule 110 would be defined by the following local rule :
%\carule{0}{1}{1}{0}{1}{1}{1}{0}

% We need one more definition yet to understand the rest of the paper :

% \begin{definition}
% A cellular automaton $\mathcal{A}$ with set of states $Q$
% is \emph{Turing-universal} if for any recursive language
% $\mathcal{L}$, there is an encoding $\varphi$ of
% $\mathcal{L}\rightarrow Q^*$ and a decoding
% $\overline{\varphi}:Q\rightarrow\{0,1\}$ such that
% $$\forall l\in\Sigma^*, l\in\mathcal{L}\Leftrightarrow
% \overline{\varphi}(F^{|\varphi(l)|}(\varphi(l)))=1$$
% Where $F$ is the global function of $\mathcal{A}$.

Finally, we call \emph{P-complete} a cellular automaton such that the
problem of predicting $F^n$ on all configurations of size $2rn+1$ is
P-complete.

Our work is motivated by the following idea: \emph{if we are able to
give a simple explicit description of $f^n$ (for arbitrary $n$), then
we can understand the behavior of the corresponding CA}.

\section{Communication complexity}
%% Les idees de communication complexity qui peuvent etre
%% interessantes pour des eca. Ce qu'on a regarde, et ce qu'on
%% aimerait regarder.
%\subsection{An introduction to communication complexity}

Communication complexity is a model introduced by A.~C.-C.~Yao in
\cite{yao79}, and designed at first for lower-bounding the amount of
communication needed in parallel programs. In this model we consider
two players, namely Alice and Bob, each with arbitrary computational
power and talking to each other to decide the value of a given
function.

For instance, let ${f:X\times Y\rightarrow Z}$ be a function taking pairs as
input. If we give first elements of pairs to Alice, and second to Bob,
the question communication complexity asks is ``how much information
do they have to communicate to each other in the worst case in order
to compute $f$ ?''.

More precisely, we define \emph{protocols}, which specify, at each
step of the communication between Alice and Bob, who speaks (Alice or
Bob), and what he says (a bit, 0 or 1), as a function of their
respective inputs.
%The theory of communication complexity studies the information
%exchanged by different actors to accomplish a common computation when
%the data is initially distributed among them. To tackle that kind of
%questions, A.C.~Yao~\cite{yao79} suggested the two-party model: two
%persons, say Alice and Bob, are asked to compute together $f(x,y)$,
%where Alice knows $x$ only and Bob knows $y$ only ($x$ and $y$
%belonging to finite sets). Moreover, they are asked to proceed in such
%a way that the cost --the total number of exchanged bits-- is minimal
%in the worst case.

This simple framework, and some of its variants we discuss in this
article, appear to us as a relevant way to study CA.
The tools of communication complexity 
suggest experiments  to test hypothesis
about properties of CA (see Section \ref{classes}).

%% Possible experiments

%\begin{definition}[One-round communication complexity]
%  A protocol ${\cal P}$ is an AB-one-round $f$-protocol if only Alice
%  is allowed to send information to Bob, and Bob is able to compute
%  the function solely on its input and the received information.  The
%  cost of the protocol $c_{AB}({\cal P})$ is the (worst case) number
%  of bits Alice needs to send. Finally, the AB-one-round communication
%  complexity of a function $f$ is $c_{AB}(f)=c_{AB}({\cal P^*})$,
%  where ${\cal P^*}$ is an AB-one-round $f$-protocol of minimum cost.
%  The BA-one-round communication complexity is defined in the same
%  way.
%\end{definition}
\begin{definition}
  A protocol $\mathcal{P}$ over domain $X\times Y$ and range $Z$ is a
  binary tree where each internal node $v$ is labeled either by a map
  $a_v:X\rightarrow\{0,1\}$ or by a map $b_v:Y\rightarrow\{0,1\}$, and
  each leaf $v$ is labeled either by a map ${A_v:X\rightarrow Z}$ or
  by a map ${B_v:Y\rightarrow Z}$.

  The \emph{value} of protocol $\mathcal{P}$ on input $(x,y)\in
  X\times Y$ is given by $A_v(x)$ (or $B_v(y)$) where $A_v$ (or $B_v$)
  is the label of the leaf reached by walking on the tree from the
  root, and walking left if $a_v(x)=0$ (or $b_v(y)=0$), and walking right
  otherwise. We say that a protocol computes a function ${f:X\times
    Y\rightarrow Z}$ if for any $(x,y)\in X\times Y$, its value on
  input $(x,y)$ is $f(x,y)$.
\end{definition}

Intuitively, each internal node specifies a bit to be communicated
either by Alice or by Bob, whereas at leaves either Alice or Bob
determines the final value of $f$ since she (or he) has received
enough information from the other.
\begin{remark}
  In our formalism, we don't ask both Alice and Bob to be able to give
  the final value. We do so to be able to consider protocols where
  communication is unidirectional (see below).
\end{remark}

\begin{definition}
  We denote by $\CC(f)$ the deterministic communication complexity of a
  function $f:X\times Y\rightarrow Z$. It is the minimal depth of a
  protocol tree computing $f$.
\end{definition}

\begin{figure}[ht]
\centering
\fbox{\includegraphics[scale=0.3]{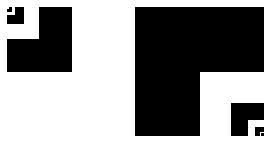}}\hspace*{30pt}
\fbox{\includegraphics[scale=0.3]{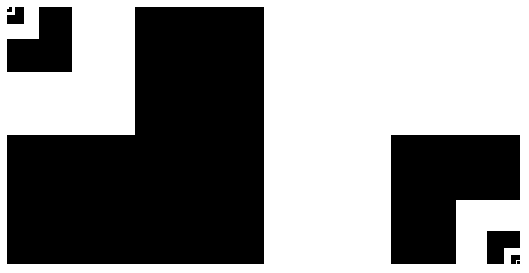}}
\caption{Matrices of rule 178, for $n=6$ (left) and $n=7$ (right)}
\label{fig178}
\end{figure}

We  study functions with the help of their associated
matrices. In such matrices, rows are indexed by elements in $X$,
columns by elements in $Y$. They are defined by $M_{i,j}=f(i,j)\in
Z$. For elementary CA, we represent the 
$n$-th iteration function of $f^n:\{0,1\}^{2n+1}\rightarrow \{0,1\}$ as
$f^n:\{0,1\}^{n} \times \{0,1\}^{n+1}\rightarrow \{0,1\}$.
For instance, Figure \ref{fig178} represents the matrix of
elementary CA rule 178, when we give $n$ bits to Alice (rows) and
$n+1$ bits to Bob (columns); i.e. when $X=\{0,1\}^n$ and
$Y=\{0,1\}^{n+1}$. We denote as $M_{178}^n$ such a matrix.

From the study in \cite{kushilevitz97}, we know that a protocol for a
function induces a partition of the matrix of this function into
monochromatic generalized rectangles (i.e. cartesian products of
subsets of $X$ and $Y$). So a lower bound for the deterministic
communication complexity of a function $\phi$ is given by
$\log_2\CP(\phi)$, where $\CP(\phi)$ stands for the \emph{partition
  number} of $\phi$, i.e. the number of rectangles needed in a minimal
partition of the matrix into monochromatic rectangles.

Moreover, we call \emph{one-round communication complexity}, denoted
by $\CCO$, the communication complexity when restricted to protocols
where only one person (Alice or Bob) can speak. Precisely, a
\emph{one-round} protocol is a tree where either all internal nodes
have labels of type $a_v$ and all leaves labels of type $B_v$ (Alice
speaking to Bob who then gives the final answer), or all internal
nodes have labels of type $b_v$ and all leaves labels of type $A_v$
(Bob speaking to Alice who gives the final answer).

\begin{definition}
  The \emph{one-round} deterministic communication complexity of a
  function $f:X\times Y\rightarrow Z$, denoted by $\CCO(f)$, is the
  minimal depth of a one-round protocol tree computing $f$.
\end{definition}

This restriction is justified by the ease of experimental measures on
the communication complexity of cellular automata it allows. More precisely,
according to Fact \ref{fact:oneway}, simply counting the number of
different rows in a matrix gives the exact one-round communication
complexity of a rule, while measuring the deterministic communication
complexity of a function implies being able to find an optimal
partition of its matrix into monochromatic rectangles.

\begin{fact}[from \cite{kushilevitz97}]
  \label{fact:oneway}
  Let $f$ be a binary function of ${2n}$ variables and
  $M_f\in\{0,1\}^{2^n\times2^n}$ its matrix representation, defined by
  $M_f(x,y)=f(xy)$ for $x,y\in\{0,1\}^n$. Let $d(M_f)$ be minimum
  between the number of different rows and the number of different
  columns in $M_f$.  We have $\CCO(f) =
  \left\lceil\log\bigl(d(M_f)\bigr)\right\rceil.$
\end{fact}

When several rounds are allowed, the communication complexity is
connected to the rank of matrices.  In fact, for an arbitrary boolean
function $f$, we have the following bounds (see~\cite{kushilevitz97}):
\[\mathrm{rank}(M_f)\geq \CC(f)\geq\log(\mathrm{rank}(M_f))\]

Moreover, the following conjecture appears in \cite{TR94} :
\begin{openproblem}
Is there a constant ${c>1}$ verifying, for any function $f$ :
\[\CC(f)\in O\bigl(\log(\mathrm{rank}(M_f))^c\bigr).\]
\end{openproblem}

Experimentally, the rank of matrices is the only parameter
we computed
in order to evaluate the multi-round communication complexity of CA.  
But it did not give tight bounds and the matrices to be
considered are exponentially large.

A theorem by J. Hromkovi\v c and G. Schnitger \cite{hromkovic97}
upper bounds the communication complexity of
Turing computations:
\begin{theorem}
  \label{turingcc}
  For a language $L\subseteq\{0,1\}^*$ and a nondeterministic TM $A$
  recognizing this language, we
  have $$T_A(n)\in\Omega\bigl(\CC(\chi_n(L))^2\bigr)$$ Where $T_A(n)$
  is the time required by $A$ to recognize $L$ and $\chi_n$ is the
  characteristic function of $L$ restricted to length $n$.
\end{theorem}
The proof uses the crossing sequence argument, introduced by Cobham
\cite{cobham64}
%    We take $c_n$ to be the nondeterministic complexity of a minimal
%    protocol on all inputs of size $n$. We are going to build a
%    protocol that has this complexity and gives evidence about the
%    time required by $A$.

%    Let $C_i$ be the $i^{th}$ crossing sequence, i.e. the sequence of
%    states in which the machine is when the head crosses a line
%    between the $i^{th}$ and the $(i+1)^{th}$ position on the Turing
%    tape. Alice first nondeterministically guesses an index
%    $j\in\{\lceil n/2\rceil-c_n/2,\ldots,\lceil n/2\rceil\}$ and a
%    possible crossing sequence $C_j$.  She sends $\lceil n/2\rceil-j$,
%    the sequence $C_j$ and the word $w_{j+1}\ldots w_{\lceil
%      n/2\rceil}$.  Then Bob checks whether $C$ is a possible crossing
%    sequence for his part of the word and accepts if there is an
%    accepting computation on $w$ with $C$ as the $i^{th}$ crossing
%    sequence. We need to send at most
%    $$\lceil n/2\rceil-i+q|C_i|+\log_2(\lceil n/2\rceil-i)$$
%    bits. Thus there will be a word on which no crossing sequence
%    between $\lceil n/2\rceil-c_n/2$ and $\lceil n/2\rceil$ may be
%    short.

%%\section{Actual results}
%%\subsection{In the general theory of cellular automata}
%% Trucs de Guillaume
%%\subsection{In elementary cellular automata}

\section{Communication Complexity in Cellular Automata}
\label{sec:ccinca}

We are interested in the sequence of iterations
${(f^n)_n}$ of the local rule of CA. So we won't consider the
communication complexity of a single function but the sequence of
complexities associated to the family ${(f^n)_n}$.

Another important point is the choice of how the input is split into 2
parts. We consider
any possible splitting into 2 connected parts and take the worst
case. Formally, given a CA local rule ${f: S^{2r+1}\rightarrow S}$, we
denote by $f_i$ (with ${0\leq i\leq 2r+1}$) the function ${f_i :
  S^i\times S^{2r+1-i}\rightarrow S}$. We also define $f_i^n$ for all
$n\geq 1$ and all $i$ with ${0\leq i\leq 2rn+1}$.

\begin{definition}
  The communication complexity $\CC(\ACA)$ of $\ACA$ is the function 
  \[n\mapsto \max_{0\leq i\leq 2rn+1}\CC(f_i^n),\] where $f$ and $r$
  are the local rule and radius of $\ACA$. We define in a similar way
  the one-round communication complexity $\CCO(\ACA)$.
\end{definition}

\begin{remark}
  This definition with arbitrary splitting of input is a slight
  modification of the definition proposed by E. Goles and I. Rapaport
  in \cite{durr04}, where the central cell is fixed, and Alice and Bob
  recieve exactly the same number of input cells.
\end{remark}

Maximal communication complexity can be reached by cellular automata.

\begin{proposition}[\cite{durr04}]
  \label{prop:existlinear}
  There is a CA $\ACA$ such that ${\CC(\ACA)\in \Omega(n)}$.
\end{proposition}

\subsection{Separation results}

One could ask whether counting the number of different rows is a
really accurate measure, and how large is the gap between the cost of
one-round protocols and the cost of protocols where several rounds are
allowed. We already know from \cite{kushilevitz97} that the gap
between one-round protocols and multi-round protocols can be
exponential. The following fact shows that we get the same exponential
gap if we restrict ourselves to functions predicting CA.

\begin{proposition}
  \label{multiway-exp-oneway}
  There exists a CA $\ACA$ such that $\CCO(\ACA)$ is an exponential
  in $\CC(\ACA)$.
  \begin{proof}[Proof sketch]
    For general functions in
    $\{0,1\}^*\times\{0,1\}^*\rightarrow\{0,1\}$, there is one
    canonical function satisfying this relation between $\CC_1$ and
    $\CC$ : consider the complete binary tree with height $h$ and
    label all its leaves and nodes with 0 or 1. The path associated to
    such a labeling is defined as follows : upon arriving on a node
    labeled with a 0 (resp. a 1), define the next node of the path as
    the root of the left (resp. right) subtree. The final value of the
    function is the label of the last node of the path (i.e. a leaf).

    In this tree, give all odd levels to Alice and even ones to
    Bob. An easy multi-round protocol solves it in communication
    complexity $h$ : in each turn, either Alice or Bob tells each
    other the label of the current node, giving to the other one the
    direction to follow (left or right) to get to the next node. It is
    a known fact from \cite{kushilevitz97} that this problem cannot be
    solved with a one-round protocol in $o(2^n)$ rounds.
    
    We describe how a CA can encode this problem on figure
    \ref{fig_multiway-exp-oneway}, in terms of signals. The top of the
    tree is encoded on the sides of the initial configurations, and
    the final values (leaves) are at the center. The squares delimit
    the levels of the tree. Clearly, odd levels are on the left, while
    even ones are on the right side of the configuration.
    
    The general behavior of this CA is to select data from the bottom
    of the tree. The green signals represent the data, the dashed ones
    represent data already selected, and the black ones are the
    selectors. All of them can carry the values 0 or 1. dotted signals
    separates the levels, transforming dashed signals into green ones.
    Black signals are selectors, they carry the values 0 (resp. 1) and
    transform into red signals carrying the value of the first
    (resp. second) green signal crossed.

    At each step, the set of ``selected'' leaves is halved by
    selections by black signals. Since these signals select the
    ``correct'' (i.e. left of right, depending on their label)
    subtree, the last leaf remaining is the actual value of this
    instance of the tree problem.
    \begin{figure}[ht!]
      \begin{center}
        \begin{tabular}{cc}
          \includegraphics{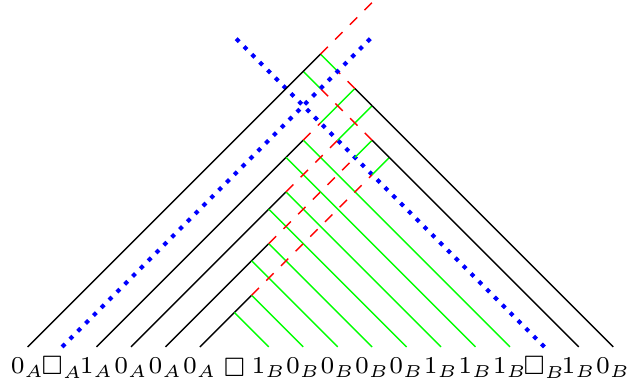} &
          \includegraphics{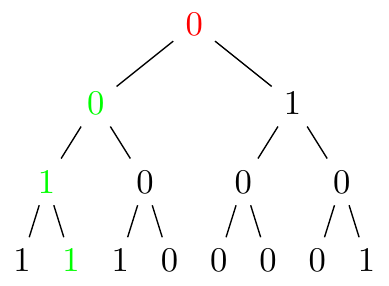}

        \end{tabular}
      \label{fig_multiway-exp-oneway}
      \caption{A cellular automaton computing the tree problem, and
        the corresponding tree}
      \end{center}
    \end{figure}
  \end{proof}
\end{proposition}

The problem with the previous proof is that we build an artificial and
complicated CA, with many states and an unclear local rule. A more
accurate question is : Are there elementary CA with a low multi-round
communication complexity, but a high lower bound for one-round
protocols? We leave this as an open problem.

\label{pcompleteness_cc}
Considering the recent results by D. Woods and T. Neary
\cite{woodsneary06}, a very natural question one could ask is the
following: What do computational properties of CA, such as
P-completeness, imply on the its communication complexity?  As shown
by the following proposition, one can build P-complete cellular
automata with arbitrarily low communication complexity.

\begin{proposition}
  For any ${k\geq 1}$, there exists a P-complete CA $\ACA$ such that
  ${\CC(\ACA) \in O(n^{1/k})}$.
  \begin{proof}[Proof sketch]
    Consider any Turing machine $\mathcal{M}$.  We construct a CA
    $\ACA$ able to simulate $\mathcal{M}$ only slowly but still in
    polynomial time: it takes $n^k$ steps of $\ACA$ to simulates $n$
    steps of $\mathcal{M}$. Hence, by a suitable choice of
    $\mathcal{M}$, $\ACA$ is P-complete.

    First it is easy to construct a CA simulating $\mathcal{M}$ in
    real time.  We encode each symbol of the tape alphabet of the
    Turing machine by a CA state, and add a ``layer'' for the head,
    with '$\rightarrow$' symbols on its left and '$\leftarrow$'
    symbols on its right. We guarantee this way that there can be only
    one head : if a '$\rightarrow$' state is adjacent to a
    '$\leftarrow$' state without head between them, we propagate an
    ``error'' state destroying everything.

    We then add a new layer to slow down the simulation: it consists
    in a single particle (we use the same trick to ensure that there
    is only one particle) moving left and right inside a marked region
    of the configuration. More precisely, it goes right until it
    reaches the end of the marked region, then it adds a marked cell
    at the end and starts to move left to reach the other end, doing
    the same thing forever. Clearly, for any cell in a finite marked
    region, seeing $n$ traversals of the particle takes $\Omega(n^2)$
    steps. Then, the idea is to authorize heads moves in the previous
    construction only at particle traversals. This way, $n$ steps of
    $\mathcal{M}$ require $n^2$ time steps of the automaton.  By
    adding another layer, one can also slow down the above particle
    with the same principle and it is not difficult to finally
    construct a CA $\ACA$ such that $n$ steps of $\mathcal{M}$ require
    $n^k$ time steps of $\ACA$.
    
    Now, the communication complexity of $\ACA$ is $O(n^{1/k})$
    because on any input of size $n$, either the ``error'' appears, or
    a correct computation of $O(n^{1/k})$ steps of $\mathcal{M}$
    occurs. Distinguishing the two cases takes only constant
    communication. Moreover, in the case of a correct computation, it
    is sufficient to determine the initial position of particles, the
    sizes of marked regions (cost $O(\log(n)$), and the initial
    position of the Turing heads as well as the $O(n^{1/k})$
    surrounding states.
  \end{proof}

\end{proposition}

\subsection{Upper bounds}
We propose here a first scheme of complexity classes in cellular
automata, based on their communication complexity. What we actually
measure is $2^{D(f)}$. This is mainly justified by experiments : the
protocols we got for cellular automata with $D(f)=2\log n$ seemed much
more sophisticated than those in $\log n$. This is also justified by
the fact that what we actually compute is either the number of
different rows or columns, or the number of rectangles in the
matrices. Thus, in the rest of this article, we will use the terms
\emph{bounded} or \emph{constant} for CA with communication complexity
bounded by a constant, \emph{linear} for CA with communication
complexity $\log n+O(1)$, \emph{quadratic} for CA with communication
complexity $2\log n+O(1)$, and so on.

In this section, we give some well-known properties of CA that induce
a bounded communication complexity.  The results below are adaptations
of ideas of~\cite{durr04} to the formalism adopted in the present
paper.

\begin{proposition}
  Let $\ACA$ be any CA of local function $f$.  If there is a function
  ${g:\setN\rightarrow\setN}$ such that $f^n$ depends on only $g(n)$
  cells, then ${\CC(\ACA)\leq g(n)/2}$.
\end{proposition}

Following the work of M.~Sablik~\cite{sablik08}, one can characterize
the set of CA having a bounded number of dependant cells
(\textit{i.e.} a bounded function $g(n)$): they are exactly these CA
which are equicontinuous in some direction (theorem 4.3
of~\cite{sablik08}). This set contains the nilpotent CA (a CA is
\emph{nilpotent} if it converges to a unique configuration from any
initial configuration, i.e. $f^n$ is a constant for any large enough
$n$).

\begin{corollary}
  If $\ACA$ is equicontinuous in some direction then ${\CCO(\ACA)}$ is
  bounded.
\end{corollary}

Another set of CA with that property is the set of linear CA.  A CA
$\ACA$ with state set $S$, radius $r$ and local global rule $G$ is
linear if there is an operator $\oplus$ such that ${(S,\oplus)}$ is a
semi-group with neutral element $e$ and for all configurations $c$ and
$c'$ we have:
\[G(c\ \overline{\oplus}\ c') = G(c)\ \overline{\oplus}\ G(c'),\] where
$\overline{\oplus}$ is the uniform extension of $\oplus$ to
configurations.

\begin{proposition}
  If $\ACA$ is linear then ${\CCO(\ACA)}$ is bounded.
\end{proposition}

The proof appears in~\cite{durr04} in a different setting. The idea is
that there is a simple one-round protocol to compute linear functions:
Alice and Bob can each compute on their own the image the function
would produce assuming the other party has only the neutral element as
input, then Alice or Bob communicate this result to the other who can
answer the final result by linearity.

\subsection{Simulation and universality}

Since the pioneering work of J.~von~Neumman \cite{neumann67},
universality in CA has received a lot of attention
(see~\cite{surveyOllinger} for a survey). Historically, the notion of
universality used for CA was more or less an adaptation of the
classical Turing-universality. Later, a stronger notion called
\emph{intrinsic universality} was proposed: a CA is intrinsically
universal if it is able to simulate any other CA. This definition
relies on a notion of simulation which is formalized below.

The base ingredient is the relation of sub-automaton.  A CA $\ACA$ is
a \emph{sub-automaton} of a CA $\ACB$, denote ${\ACA\sac\ACB}$, if
there is an injective map $\iota$ from $\alphA$ to $\alphB$ such that
${\overline{\iota}\circ\globA=\globB\circ \overline{\iota}}$, where
${\overline{\iota}:\alphA^\ZZ\rightarrow \alphB^\ZZ}$ denotes the
uniform extension of $\iota$.

A CA $\ACA$ simulates a CA $\ACB$ if some \emph{rescaling} of $\ACA$
is a sub-automaton of some \emph{rescaling} of $\ACB$. The ingredients
of the rescalings are simple: packing cells into blocs, iterating the
rule and composing with a translation. Formally, given any state set
$Q$ and any $m\geq 1$, we define the bijective packing map ${\bloc{m}:
  Q^\ZZ\rightarrow \bigl(Q^m\bigr)^\ZZ}$ by:
\[\forall z\in\ZZ : \bigl(\bloc{m}(c)\bigr)(z) = \bigl(c(mz),\ldots,c(mz+m-1)\bigr)\]
for all ${c\in Q^\ZZ}$. The rescaling $\grp{\ACA}{m,t,z}$ of $\ACA$ by
parameters $m$ (packing), ${t\geq 1}$ (iterating) and ${z\in\ZZ}$
(shifting) is the CA of state set $Q^m$ and global rule:
\[\bloc{m} \circ \sigma_z\circ \globA^t \circ \debloc{m}.\]
With these definitions, we say that $\ACA$ simulates $\ACB$, denoted
${\ACA\simu\ACB}$, if there are rescaling parameters $m_1$, $m_2$,
$t_1$, $t_2$, $z_1$ and $z_2$ such that
${\grp{\ACA}{m_1,t_1,z_1}\sac\grp{\ACB}{m_2,t_2,z_2}}$.

We can now naturally define the notion of universality associated to
this simulation relation.

\begin{definition}
  $\ACA$ is \emph{intrinsically universal} if for all $\ACB$ it holds
  ${\ACB\simu\ACA}$.
\end{definition}

This definition of universality may seem very resctrictive. In fact,
many so-called universal CA (\textit{i.e.} Turing-universal CA) are
also intrinsically universal (see~\cite{surveyOllinger}
and~\cite{liferokadur} for the particular case of Game of Life),
although there is still a gap for one-dimensional CA (the elementary
CA 110 is Turing-universal and no elementary CA is known to be
intrinsically universal). Moreover, intrinsic universality appears to
be very common in some classes of CA (see~\cite{Theyssier05}).

But, most importantly, by completely formalizing\footnote{There is
  actually no consensus on the formal definition of
  Turing-universality in CA (see~\cite{liferokadur} for a discussion
  about encoding/decoding problems).} the notion of universality, we
facilitate the proof of negative results.

We are going to show that the tool of communication complexity is
precisely a good candidate to obtain negative results. The idea is
simple: if $\ACA$ simulates $\ACB$ then the communication complexity
of $\ACA$ must be 'greater' than the communication complexity of
$\ACB$.

More precisely, we consider the following relation of comparison
between functions from $\setN$ to $\setN$:
\[\phi_1\moindre \phi_2\iff\exists \alpha,\beta,\gamma\geq 1,\forall n\in\setN:
\phi_1(\alpha n)\leq\beta\phi_2(\gamma n).\]

\begin{proposition}
  If $\ACA\simu\ACB$ then ${\CC(\ACA)\moindre\CC(\ACB)}$.
\end{proposition}
\begin{proof}[Proof sketch]
  We consider successively each ingredient involved in the simulation
  relation: 
  \begin{description} \item[Sub-automaton: ] if ${\ACA\sac\ACB}$ then
    each valid protocol to compute iterations of $\ACB$ is also a
    valid protocol to compute iterations of $\ACA$ (up to state
    renaming).  \item[Iterating: ] the complexity function of $\ACA^t$
    is ${n\mapsto\phi(t\cdot n)}$ if $\phi$ is the complexity function
    of $\ACA$.  \item[Shifting: ] this operation only affects the
    splitting of inputs. Since we always take in each case the
    splitting of maximum complexity, this has no influence on the
    final complexity function.  \item[Packing: ] let $\ACA$ be CA with
    local rule $f$ and states set $S$. Consider any sequence of valid
    protocols $(P_j)$, one for each splitting of inputs of $f^n$, and
    denote by ${h:(S^m)^i\times(S^m)^{k-i}\rightarrow S^m}$ some
    splitting of the $n$th iteration of the local rule of
    ${\grp{\ACA}{m,1,0}}$. By definition of packing map $\bloc{m}$, a
    valid protocol for $h$ is deduced by simultaneous application of
    protocols ${P_j,\ldots, P_{j+m-1}}$ (for a suitable choice of
    $j$), each being used to determined one component of the resulting
    value of $h$ which belongs to $S^m$. It follows that ${\CC(h)\leq
      m\cdot\CC(f^n)}$.  
  \end{description} 
  Therefore we have:
  ${\CC(\ACA)\moindre\CC\left(\grp{\ACA}{m,t,z}\right)}$,
  ${\CC\left(\grp{\ACA}{m,t,z}\right)\moindre\CC(\ACA)}$ and if
  ${\ACA\sac\ACB}$ then ${\CC(\ACA)\moindre\CC(\ACB)}$.
\end{proof}

From Proposition~\ref{prop:existlinear}, we derive the following
necessary condition for intrinsic universality. It is one of the main
motivations to study communication complexity of CA,
both theoretically and experimentally.

\begin{corollary}
  If $\ACA$ is intrinsically universal then ${\CC(\ACA)\in \Omega(n)}$.
\end{corollary}

\section{The one-round communication complexity of ECA}
\label{classes}

In this section we concentrate on elementary cellular automata (ECA)
: dimension one, two states, and radius $r=1$. And we split the input as
follows : $f:\{0,1\}^{n} \times \{0,1\}^{n+1} \rightarrow \{0,1\}$.
Since any ECA has the same (one-round) communication complexity as its
reflex and its conjugate, we propose here a classification of the 88
nonisomorphic ECA. Since we only consider one-round communication
complexity here, Fact~\ref{fact:oneway} allows us to consider matrices
associated to functions and study the number of their different rows
or columns.

Therefore, for the sake of clarity, the name we give to classes of ECA
is related to the number of different rows and columns (instead of the
one-round communication complexity, which is the logarithm of the
previous).

\subsection{Bounded (by a constant)}
As shown above, several results allow us to bound the (one-round)
communication complexity of many CA. 

The ECA proved to be in this class are the following 44 ones:
%\bigskip
0, 1, 2, 3, 4, 5, 7, 8, 10, 12, 13, 15, 19, 24, 27, 28, 29, 32, 34,
36, 38, 42, 46, 51, 60, 72, 76, 78, 90, 105, 108, 128, 130, 136, 138,
140, 150, 156, 160, 162, 170, 172, 200, 204 
(and all their reflexes, conjugates, and reflex-conjugates).

\subsection{Linear}

Consider for instance rule 178, which has been studied recently by
D. Regnault~\cite{regnault08} using percolation theory. The author
considered the case where each cell has an independent probability
$\rho$ to be updated in each step. He studied Rule 178 because it
``exhibited rich behavior such as phase transition''. Despite its
complexity, this CA was amenable to formal analysis: the proofs were
based on a coupling between its space-time diagram and oriented
percolation on a graph.

It is not difficult, using the methods of \cite{kushilevitz97}, to
prove that the communication complexity of CA 178 grows as
$\Theta(n)$. Notice that in order to get such a result we must find, on
one hand, a communication protocol (upper bound) of complexity
$\approx \log(n)$ and, on the other hand, to exhibit a ``fooling set''
(i.e. a set $C$ of configurations such that for any couple $(x,y)$ of
configurations of $C$, $x$ and $y$ are necessarily in two distinct
monochromatic rectangles) of size in $\Omega(n)$.

\label{rule178}

The ECA Rule 178  is given by the following local
rule : \carule{0}{1}{0}{0}{1}{1}{0}{1}

There is a very simple protocol ${\cal P}$  
in $\log n+1$ bits for it: if we call
$c$ the value of the central cell at the beginning (Bob knows it),
then Bob sends the length of the longest string of cells with value
$c$, starting from the left of his part, to Alice.

\begin{proposition}
Protocol ${\cal P}$ is correct for ECA Rule 178.
\begin{proof}
  First remark that configurations $01$ and $10$ map to each other for
  any values of their right or left neighbour (which we can see in
  figure \ref{178-dep} where undetermined cells are represented in
  gray), and thus stay stable.
  \begin{figure}[ht]
    \centering
    \includegraphics[scale=0.8]{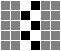}
    \caption{Evolution of $01$ and $10$ for rule 178}
    \label{178-dep}
  \end{figure}
  So once Alice knows where the first $01$ or $10$ occurs, she can
  assume w.l.g. that the rest of Bob's part are only zeros (the
  final result is the same). But then, she also knows the beginning of
  Bob's part, so she can compute the final result of Rule 178.
\end{proof}
\end{proposition}

\begin{proposition}
Protocol ${\cal P}$ is optimal even as a multi-round protocol.
\begin{proof}
  To show this, we use the results of \cite{kushilevitz97} and exhibit
  a fooling set. Let
  $$C=\{(0^{n-2k-1}10^{2k},c^{2k}\overline{c}c^{n-2k})|
  0\leq 2k\leq n-1, c\in\{0,1\}\}$$ 
  
  First remark that the result of Rule 178 on configurations of the
  form
  $$0^{n-2k-1}10^{2k}c^{2k}\overline{c}c^{n-2k}$$
  is always $n\mod 2$, while for any $i\neq j$, the result of
  $0^{n-2i-1}10^{2i}c^{2j}\overline{c}c^{n-2j}$ is $n+1\mod 2$.

  For the case $c=1$, this is shown by our previous remark on stable
  configuration. For $c=0$, this is a simple remark on the space-time
  diagrams of rule 178.

  Then $|C|=2\cdot\lfloor n/2\rfloor$, and thus no deterministic
  protocol, even multiround, could predict rule 178 in less than $\log
  n+1$ rounds.
\end{proof}
\end{proposition}

\begin{remark}
  The same argument can be used for rule 50 (and thus also 179).
\end{remark}

We believe that the linearity of Rule 178 and the fact that it is
amenable to other types analysis is not a coincidence.

\subsection{Quadratic}
As soon as we move up in our hierarchy the underlying protocols become
rather sophisticated. In fact, for Rule 218, we prove in~\cite{GLR08}
that if $c=0$ then Alice needs to send 2 positions of her string (2
times $\log(n)$ bits).  The difference in the difficulty between
sending 1 position ($\Theta(n)$ behavior) and 2 positions
($\Theta(n^2)$ behavior) is huge.

We encountered Rule 218 when trying to find a (kind of)
double-quiescent palindrome-recognizer.  Despite the fact that it
belongs to class II (according to Wolfram's classification), it mimics
Rule 90 (class III) for very particular initial configurations.

Behind the following ``proofs'' there are lots of lemmas that we are
not even stating.  Therefore, the purpose here is just to give an idea
of how we proceed.  We are considering the case when the central cell
is 0. We write $f$ instead of $f_{218}$.

\begin{definition}
  We say that a word in $\{0,1\}^*$ is additive if the 1s are isolated
  and every consecutive couple of 1s is separated by an odd number of
  0s.
\end{definition}

\begin{notation}
  Let $\alpha$ be the maximum index $i$ for which $x_i\ldots x_10$ is
  additive. Let $\beta$ be the maximum index $j$ for which $0y_1
  \ldots y_j$ is additive. Let $x'=x_{\alpha}\ldots x_1 \in
  \{0,1\}^{\alpha}$ and $y'=y_1 \ldots y_{\beta} \in \{0,1\}^{\beta}$.
\end{notation}

\begin{notation}
  Let $l$ be the minimum index $i$ for which $x_i=1$. If such index
  does not exist we define $l=0$. Let $r$ be the minimum index $j$ for
  which $y_j=1$. If such index does not exist we define $r=0$.
\end{notation}

\begin{proposition}
\label{prop:prot0}
There exists a one-round $f$-protocol ${\cal P}_0$ with cost
$2\lceil\log(n)\rceil+1$.
\end{proposition}

\begin{proof}
  Recall the Alice knows $x$ and Bob knows $y$.  ${\cal P}_0$ goes as
  follows. Alice sends to Bob $\alpha$, $l$, and
  $a=f^{\alpha}(x',0,0^{\alpha})$.  The number of bits is therefore
  $2\lceil\log(n)\rceil+1$.

  If $l=0$ then Bob knows (by definition of $l$) that $x=0^n$ and he
  outputs $f^n(0^n,0,y)$. If $r=0$ his output depends on $\alpha$.  If
  $\alpha=n$ he outputs $a$ and if $\alpha <n$ he outputs 1.  We can
  assume now that neither $l$ nor $r$ are 0.  The way Bob proceeds
  depends mainly on the parity of $|l+r-1|$.

\medskip

\noindent {\bf Case $|l+r-1|$ is odd.} If $|\alpha-\beta| \geq 1$ Bob
outputs 1. If $\alpha=\beta=k$ he outputs $a+f^k(0^k,0,y')$.

\medskip

\noindent {\bf Case $|l+r-1|$ is even.} Bob compares $r$ with $l$. If
$l \geq r-1$ then Bob outputs $f^n(1^{n-l+1}0^{l-1},0,y)$ if $l \geq
r+3$ and 1 otherwise. If $l \leq r-3$ then he outputs
$a=f^{\alpha}(x',0,0^{\alpha})$ if $r=\alpha+1$ and 1 otherwise.
\end{proof}

Now we exhibit lower bounds for the number of
different rows of the corresponding matrix. If these bounds appear to be tight
then, from Fact~\ref{fact:oneway}, they can be used for proving the
optimality of our protocol.

\begin{proposition}
  \label{prop:opt0} The cost of any one-round $f$-protocol is at least
  $2\lceil\log(n)\rceil-5$.
\end{proposition}

\begin{proof}
  Consider the following subsets of $\{0,1\}^n$. First,
  $S_3=\{1^{n-3}000\}$. Also, 
  $$S_5=\{1^{n-5}00000,1^{n-5}01000\}.$$
  In general, for every $k \geq 2$ such that $2k+1 \leq n$, we define

  $$S_{2k+1}=\{1^{n-2k-1}0^{2k+1}\} \cup \{1^{n-2k-1}0^a10^b|
  \mbox{ $a$ odd, $b$ odd, $b \geq 3$, $a+b=2k$}\}.$$

  Let $x_n \ldots x_1 \in S_{2k+1}$ and $\tilde x_n \ldots \tilde x_1
  \in S_{2\tilde{k}+1}$ with $k \neq \tilde{k}$.  It follows that the
  rows of $M_f^{c,n}$ indexed by $x_n \ldots x_1$ and $\tilde x_n
  \ldots \tilde x_1$ are different.

  Let $x=x_n \ldots x_1$, $\tilde{x}= \tilde x_n \ldots \tilde x_1 \in
  S_{2k+1}$ with $x \neq \tilde x$. It follows that there exists
  $y=y_1 \ldots y_n \in \{0,1\}^n$ such that $f^n(x,0,y) \neq
  f^n(\tilde x,0,y)$.
\end{proof}

\subsection{Non-polynomial}
Our experiments suggested the existence of (at least) two subclasses
of this class of ``hard'' ECA.
\begin{itemize}
\item Automata with a high one-round communication complexity but a
  low matrix rank (suggesting a low multi-round communication
  complexity), meaning they are easy to predict with several actors
  and a protocol between them, but the exact influence of each cell of
  the initial state is hard to determine. We do not know whether this
  class really exists among ECA, but our experiments suggest that rule
  30 may be a candidate.

\item Automata that are ``intrinsically hard'', meaning that they do
  not have a deterministic protocol in the previous classes.
\end{itemize}

%%\subsection{Study of four specific rules}
%%\input{rules}
\section{Conclusion and perspectives}
\focus{Input splitting.} When defining the communication complexity in
CA we consider the worst case for splitting the inputs
for each $n$. We believe that the sequence ${(s_n)_n}$ of
such worst-case splittings is meaningful and raises several
interesting questions: Is $s_n$ unique for each $n$? If it is the
case, what is the function ${n\mapsto s_n}$? Is it linear, thus
showing a direction of maximal 'information exchange' along time? What
is the meaning of such a direction?

\focus{Higher dimensional CA and multi-party protocols.} We focused
our study on the model where Alice and Bob need to communicate to
predict a given CA. There are also other models of
protocols with $k$ players, but the difficulty of experimentation
would probably not be the same. A greater number of players seems more
natural for dimension $2$ or more, since we can partition the set of
dependant cells into adjacent regions. But the two-player framework
could also be applied to higher dimensional CA.

\focus{Nondeterministic protocols.} A possible generalization of our
definitions of protocols is to allow Alice and Bob to take
nondeterministic steps in the protocol tree. This gives us other
interesting tools and measures, for instance the notion of a
\emph{cover} of a matrix, which seems linked to circuits. We can find
in \cite{kushilevitz97} a link between nondeterministic protocols and
the minimal number of rectangles needed to cover a matrix with
possible intersections between rectangles.

\focus{Probabilistic protocols.} Another relevant generalization of
communication complexity for the study of CA is
randomized complexity, where errors are allowed. In this model, Alice
and Bob are allowed to toss a coin before communicating (see
\cite{kremernr01} regarding one-round randomized complexity and
\cite{NisanW93} for many-round). Allowing randmoness just changes the
notion of complexity and can be applied to deterministic CA, but it
may make sense to use this framework for stochastic CA (see for
instance \cite{regnault08}).

\bibliographystyle{eptcs}%\bibliographystyle{elsart-num-sort}

\end{document}